\documentclass[aps,prd,reprint,twocolumn,superscriptaddress,showpacs]{revtex4-2}

\usepackage{graphicx}
\usepackage{mathrsfs}
\usepackage{bm}
\usepackage{amsmath}
\usepackage{dcolumn}
\usepackage{epstopdf}
\usepackage{dsfont}
\usepackage{amssymb}
\usepackage{tabularx}
\usepackage{array}
\usepackage{float}
\usepackage{color}
\usepackage{epstopdf}
\usepackage{mathrsfs}
\usepackage[colorlinks, linkcolor=blue,anchorcolor=blue,citecolor=blue,urlcolor=blue]{hyperref}
\usepackage{multirow}

\begin{document}

\title{Valley-dependent electronic properties in two-dimensional altermagnetic iron-based transition metal chalcogenides}
	
	\author{An-Dong Fan}
	\affiliation{School of Physics, Northwest University, Xi'an 710127, China}
	\affiliation{Shaanxi Key Laboratory for Theoretical Physics Frontiers, Xi'an 710127, China}
	
	\author{Yong-Kun Wang}
	\affiliation{School of Physics, Northwest University, Xi'an 710127, China}
	\affiliation{Shaanxi Key Laboratory for Theoretical Physics Frontiers, Xi'an 710127, China}
	
	\author{Jin-Yang Li}
	\affiliation{School of Physics, Northwest University, Xi'an 710127, China}
	\affiliation{Shaanxi Key Laboratory for Theoretical Physics Frontiers, Xi'an 710127, China}
	
	\author{Si Li}
	\email{sili@nwu.edu.cn}
	\affiliation{School of Physics, Northwest University, Xi'an 710127, China}
	\affiliation{Shaanxi Key Laboratory for Theoretical Physics Frontiers, Xi'an 710127, China}
	
\begin{abstract}
Altermagnets represent a newly identified third class of collinear magnets and have recently emerged as a focal point in condensed matter physics. In this work, through first-principles calculations and theoretical analysis, we identify monolayer Fe$_2$MoX$_4$ (X = S, Se, Te) and Fe$_2$WTe$_4$, a class of iron-based transition metal chalcogenides, as promising altermagnetic materials. These systems are found to be semiconductors exhibiting spin splitting in their nonrelativistic band structures, indicative of intrinsic altermagnetic ordering.
Remarkably, their valence bands feature a pair of valleys at the time-reversal-invariant momenta  X and Y points. Unlike conventional valley systems, these valleys are related by crystal symmetries rather than time-reversal symmetry. We investigate valley-dependent physical phenomena in these materials, including Berry curvature and optical circular dichroism, revealing strong valley-contrasting behavior.
Furthermore, we investigate the effect of uniaxial strain and show that it effectively lifts the valley degeneracy, resulting in pronounced valley polarization. Under hole doping, this strain-induced asymmetry gives rise to a piezomagnetic response. We also explore the generation of anisotropic noncollinear spin currents in these systems, expanding the scope of their spin-related functionalities.
Our findings unveil rich valley physics in monolayer Fe$_2$MoX$_4$ (X = S, Se, Te) and Fe$_2$WTe$_4$, highlighting their significant potential for applications in valleytronics, spintronics, and multifunctional nanoelectronic devices.
		
\end{abstract}
	
\maketitle
\section{Introduction}
Altermagnets (AMs), a recently identified magnetic phase, exhibit compensated magnetic order in real space yet host nonrelativistic spin-splitting in reciprocal space, attracting intense interest in condensed matter physics~\cite{vsmejkal2022beyond,vsmejkal2022emerging,bai2024altermagnetism,fender2025altermagnetism}. A hallmark of altermagnetism lies in its real-space spin sublattice configuration, where opposite spins are related by rotational symmetry rather than inversion or translation. This distinctive structural feature gives rise to a range of extraordinary physical properties, including the anomalous Hall effect~\cite{feng2022anomalous}, unique spin current generation~\cite{bai2022observation,karube2022observation,gonzalez2021efficient}, giant tunneling magnetoresistance~\cite{shao2021spin,vsmejkal2022giant}, spin Seebeck and spin Nernst effects~\cite{cui2023efficient}, Andreev reflection~\cite{sun2023andreev}, and topological superconductivity~\cite{ahn2019antiferromagnetism,zhu2023topological,li2023majorana},  ferroelectric and antiferroelectric~\cite{gu2025ferroelectric,duan2025antiferroelectric,vsmejkal2024altermagnetic}.
To date, numerous altermagnetic materials have been theoretically predicted and experimentally confirmed, including RuO$_2$~\cite{berlijn2017itinerant,zhu2019anomalous}, MnTe~\cite{gonzalez2023spontaneous,krempasky2024altermagnetic}, MnTe$_2$~\cite{zhu2024observation}, CrSb~\cite{li2024topological,lu2024observation,ding2024large,zhou2025manipulation,yang2025three}, Rb$_{1-\delta}$V$_2$Te$_2$O~\cite{zhang2024crystal}, and KV$_2$Se$_2$O~\cite{jiang2024discovery}. 

Altermagnetism, with its distinctive spin-splitting and spin-momentum locking characteristics, is catalyzing a paradigm shift in valleytronics research~\cite{schaibley2016valleytronics,vitale2018valleytronics,ma2021multifunctional}. The core concept of valleytronics lies in harnessing the ``valley" degree of freedom—multiple energy extrema in a semiconductor's Brillouin zone (BZ)—as a new information carrier, analogous to charge and spin~\cite{rycerz2007valley,gunawan2006valley,xiao2007valley,yao2008valley,xiao2012coupled,cai2013magnetic}. Prototypical platforms like graphene and transition metal dichalcogenides feature valleys at the $K$ and $K'$ points, whose degeneracy is protected by time-reversal symmetry ($T$). Consequently, achieving the crucial step of valley polarization in these systems has traditionally relied on external symmetry-breaking stimuli, such as magnetic fields~\cite{cai2013magnetic,li2014valley,aivazian2015magnetic,srivastava2015valley,macneill2015breaking,qi2015giant,jiang2017zeeman} or optical pumping with circularly polarized light~\cite{mak2012control,zeng2012valley,cao2012valley,hsu2015optically,mak2018light}.
Altermagnets fundamentally challenge this established paradigm. In these materials, valley degeneracies are governed by crystal symmetries rather than time-reversal symmetry~\cite{vsmejkal2022beyond,vsmejkal2022emerging,li2024strain}. This critical distinction liberates valley control from the constraints of time-reversal-breaking fields and opens up unprecedented pathways for manipulation. Instead of magnetism or light, valley polarization in altermagnets can potentially be engineered through more versatile means like mechanical strain~\cite{ma2021multifunctional,zhu2023multipiezo,li2024strain} or gate-tunable electric fields~\cite{zhang2024predictable}, heralding a new era of electrically and mechanically controlled valleytronic devices.
Despite significant theoretical advances, the number of experimentally confirmed 2D altermagnetic valleytronic materials remains severely limited. This materials gap forms a major obstacle to advancing valleytronics, a promising avenue for future spintronic and quantum devices. Consequently, identifying and investigating new 2D altermagnets with robust valley functionalities is now a primary research imperative.

In this work, based on first-principles calculations and theoretical analysis, we systematically investigate the valley-related physical properties of two-dimensional (2D) altermagnetic iron-based transition metal chalcogenides, Fe$_2$MoX$_4$ (X = S, Se, Te) and Fe$_2$WTe$_4$. We demonstrate that monolayer Fe$_2$MoX$_4$ (X = S, Se, Te) and Fe$_2$WTe$_4$ possess an altermagnetic ground state and exhibit substantial spin splitting even in the absence of spin–orbit coupling (SOC). Moreover, their electronic band structures exhibit semiconducting behavior with pronounced valley features at the time-reversal-invariant momenta (TRIM) X and Y points of the valence band. These valleys are energetically degenerate but carry opposite spin polarization, a consequence of the $\{C_{2}||S_{4z}^+\}$ symmetry.
We further explore valley-related physical phenomena, including Berry curvature and optical circular dichroism. We reveal that external uniaxial strain can lift the valley degeneracy, leading to valley polarization and the emergence of piezomagnetism under hole doping. In addition, we investigate the emergence of anisotropic noncollinear spin currents in these systems. This study uncovers a rich array of valley-dependent phenomena and highlights the strong potential of these altermagnetic monolayers for future applications in valleytronics, spintronics, and nanoelectronics.

\section{FIRST-PRINCIPLES METHODS}
First-principles calculations were carried out within the framework of density functional theory (DFT) using the Vienna ab initio Simulation Package (VASP)\cite{Kresse1994,Kresse1996}. The exchange-correlation potential was treated using the generalized gradient approximation (GGA) in the Perdew–Burke–Ernzerhof (PBE) form\cite{PBE}. A plane-wave energy cutoff of 600 eV was employed, and the BZ was sampled using a $\Gamma$-centered $12 \times 12 \times 1$ $k$-mesh. The convergence criteria for total energy and atomic forces were set to $10^{-7}$ eV and 0.005 eV/Å, respectively. To eliminate spurious interactions between periodic images, a vacuum layer of 20 Å was introduced along the out-of-plane direction.
The on-site Coulomb interaction for Fe 3$d$ orbitals was treated using the DFT+$U$ approach~\cite{Anisimov1991,dudarev1998}, with an effective $U$ value of 4 eV adopted for Fe, consistent with previous studies~\cite{jain2011formation}. Phonon spectra were calculated within the framework of density functional perturbation theory (DFPT) using a $3 \times 3 \times 1$ supercell, as implemented in the PHONOPY package~\cite{togo2015first}. The ab initio molecular dynamics (AIMD) simulations were performed using a $2 \times 2 \times 1$ supercell.
Spin-resolved charge conductivities were computed based on Boltzmann transport theory within the constant relaxation time approximation, using the Wannier90 package~\cite{Marzari1997,Souza2001,pizzi2014boltzwann}. An electronic temperature of 300 K and a relaxation time of 10 fs were used. A dense $300 \times 300 \times 1$ $k$-mesh was employed for BZ integration. To eliminate the vacuum contribution and ensure that the resulting units are consistent with those used for conventional three-dimensional bulk systems, we adopt a well-established scaling procedure by multiplying the supercell results by a factor of $L_z / d_{\mathrm{eff}}$. Here, $L_z$ denotes the lattice constant along the $z$ -axis (including the vacuum region), and $d_{\mathrm{eff}}$ represents the effective thickness of the system. The Berry curvature and circular dichroism were calculated using the VASPBERRY~\cite{kim2022circular}.
	
\section{RESULTS}
	
\subsection{Crystal structure and magnetism}
	
\begin{figure}[htb]
		\includegraphics[width=8.5cm]{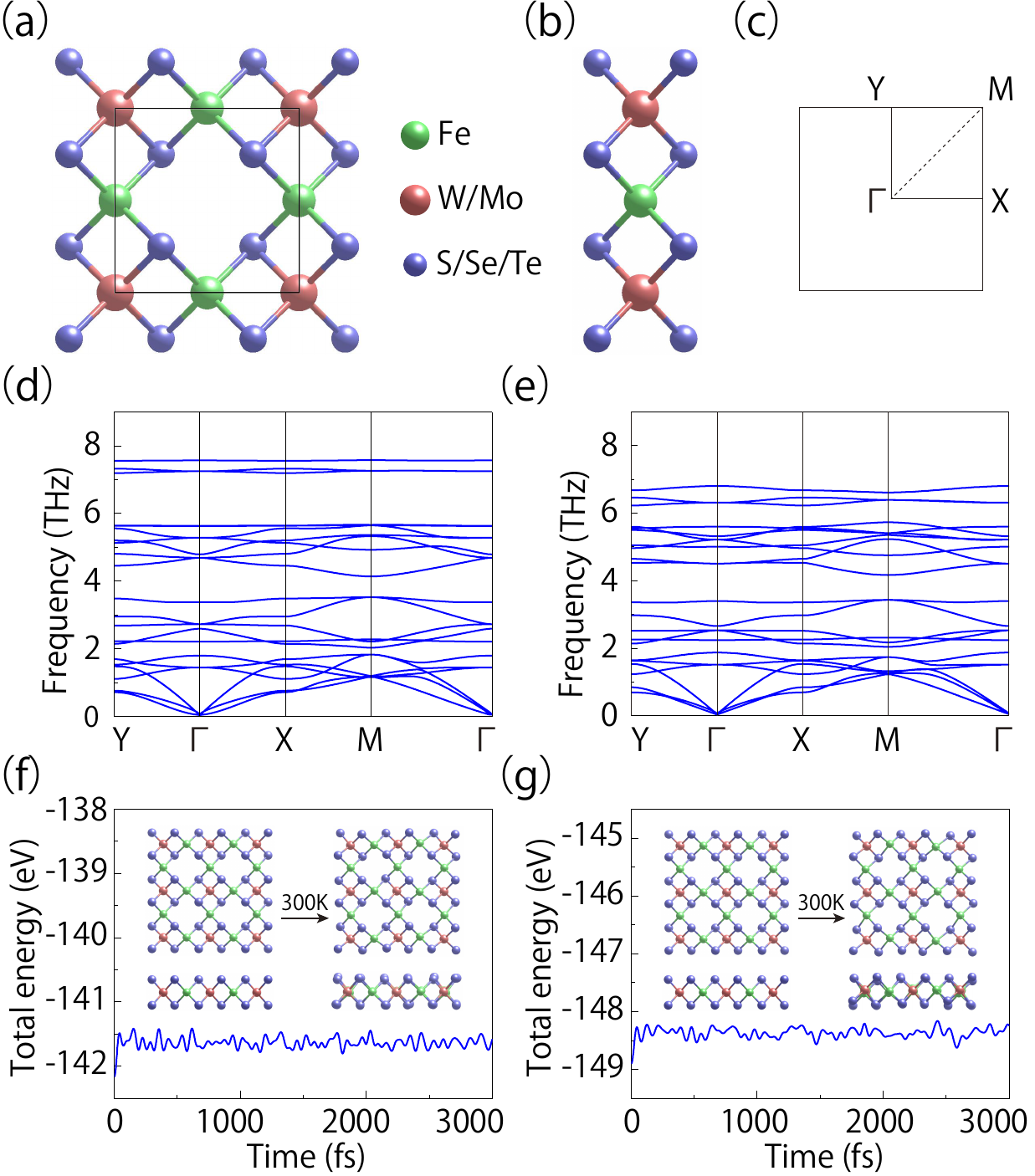}
		\caption{ (a) Top view and (b) side view of the crystal structure of the studied monolayer material. The primitive cell is shown by the solid line in (a). (c) BZ with the high-symmetry points labeled. The calculated phonon spectra of monolayer (d) Fe$_2$MoTe$_4$ and (e) Fe$_2$WTe$_4$, and the AIMD simulation results for (f) Fe$_2$MoTe$_4$ and (g) Fe$_2$WTe$_4$.
			\label{fig1}}
\end{figure}
	
\begin{figure}[htb]
	\includegraphics[width=8.5cm]{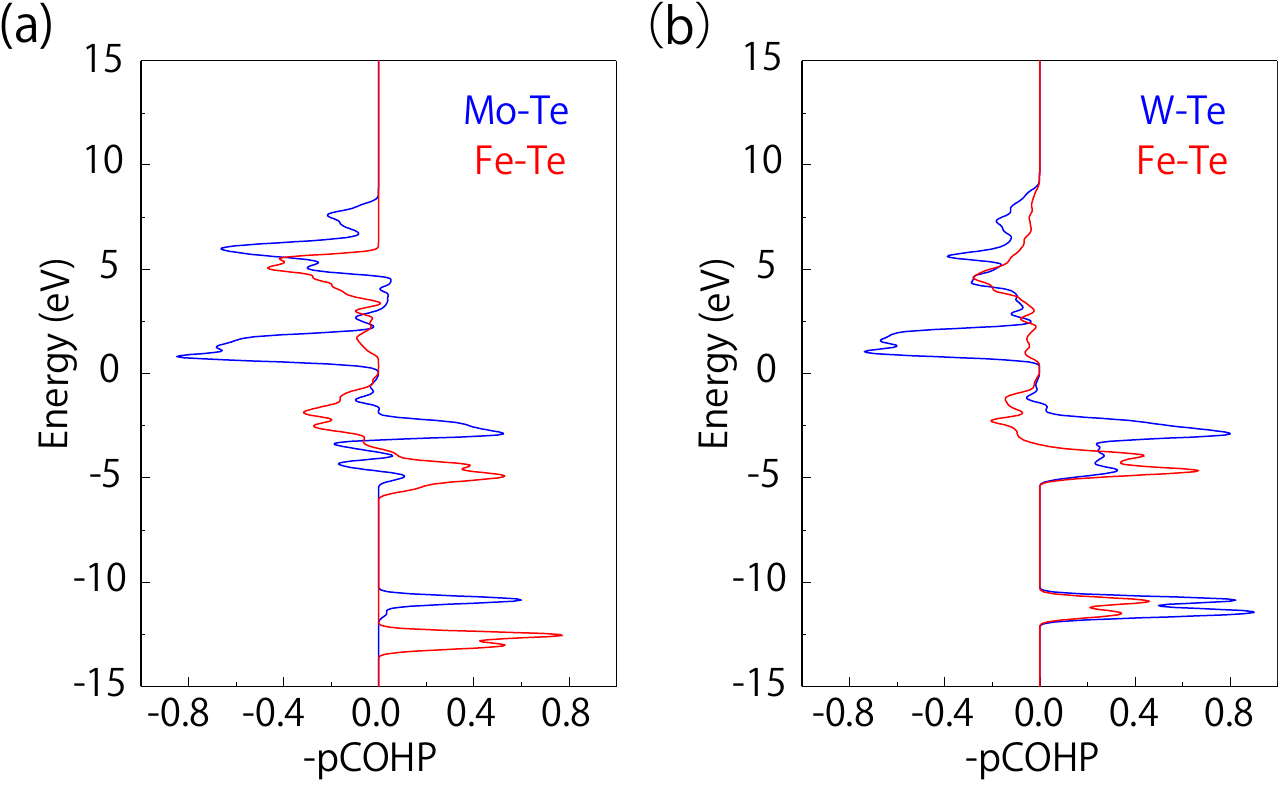}
	\caption{ The pCOHP analysis of the nearest-neighbor and next-nearest-neighbor interactions in monolayer (a) Fe$_2$MoTe$_4$ and (b) Fe$_2$WTe$_4$.
		\label{fig2}}
\end{figure}
	
\begin{figure*}[htb]
		\includegraphics[width=17cm]{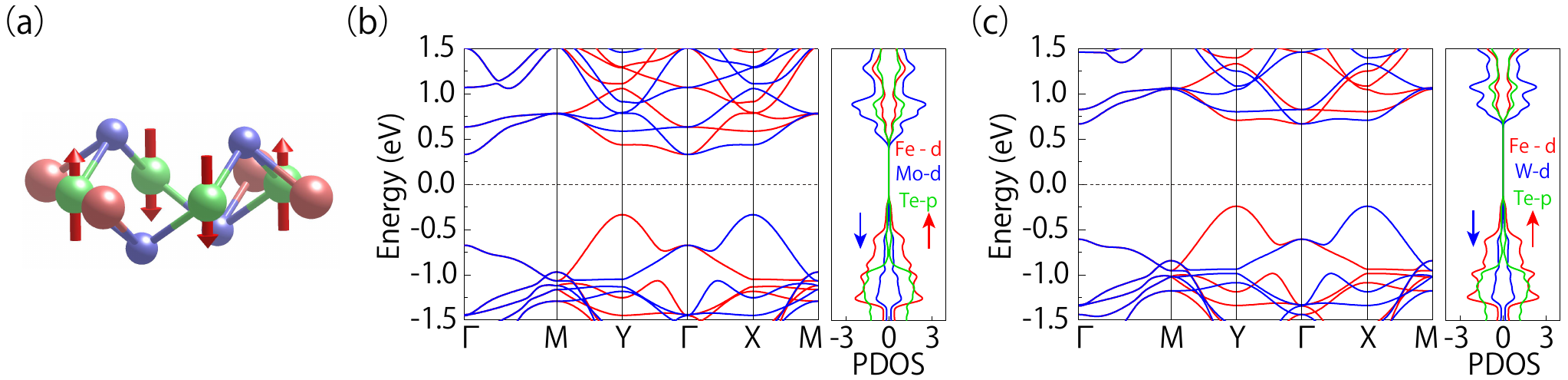}
		\caption{(a) Illustration of the altermagnetic ground state of monolayer Fe$_2$MoTe$_4$ and Fe$_2$WTe$_4$. (b) and (c) Band structures and projected density of states (PDOS) of monolayer Fe$_2$MoTe$_4$ and Fe$_2$WTe$_4$ without SOC, respectively. Red (blue) denotes spin-up (spin-down) bands.
			\label{fig3}}
\end{figure*}

Monolayer Fe$_2$MoX$_4$ (X = S, Se, Te) and Fe$_2$WTe$_4$ adopt a square lattice structure belonging to the space group $P\overline{4}2m$ (No. 111), characterized by symmetry operations generated by $S_{4z}^+$ and $C_{2x}$. As illustrated in Figs.~\ref{fig1}(a) and \ref{fig1}(b), these monolayers exhibit a trilayer architecture, in which the Fe/Mo(W) atomic layer is sandwiched between two chalcogen (X) layers. The corresponding BZ and high-symmetry points are shown in Fig.~\ref{fig1}(c). The optimized lattice constants are as follows: $a = b = 5.547$ Å for Fe$_2$MoS$_4$, $a = b = 5.698$ Å for Fe$_2$MoSe$_4$, $a = b = 5.935$ Å for Fe$_2$MoTe$_4$, and $a = b = 5.925$ Å for Fe$_2$WTe$_4$. Given the similarities in structural and electronic properties among these compounds, we primarily focus on Fe$_2$MoTe$_4$ and Fe$_2$WTe$_4$ in the main text, with the corresponding results for Fe$_2$MoS$_4$ and Fe$_2$MoSe$_4$ provided in the Supplemental Material (SM)~\cite{SM}.

We first examine the stability of monolayer Fe$_2$MoTe$_4$ and Fe$_2$WTe$_4$. To evaluate their dynamical stability, we performed phonon spectrum calculations, shown in Figs.~\ref{fig1}(d) and ~\ref{fig1}(e). The absence of imaginary frequencies across the spectra confirms that both structures are dynamically stable.  To further evaluate the thermal stability, particularly for potential room-temperature applications, we performed AIMD simulations using a $2 \times 2 \times 1$
 supercell. As shown in Figs ~\ref{fig1}(f) and ~\ref{fig1}(g), after 3000 fs of simulation at 300 K, the structures exhibited only minor thermally induced fluctuations. Importantly, no bond breaking or structural reconstruction was observed, confirming that these monolayer materials maintain excellent thermal stability under ambient conditions. Additionally, these monolayer structures are inspired by experimentally synthesized layered compounds, such as Cu$_2$MX$_4$ and Ag$_2$WS$_4$~\cite{pruss1993new,crossland2005synthesis,lin2019recent,balu2022controlled,zhan2018low}, indicating their strong potential for experimental realization. We also performed projected crystal orbital Hamiltonian population (pCOHP)~\cite{deringer2011crystal} analyses on the Mo–Te, W–Te, and Fe–Te bonds in the proposed structures of Fe$_2$MoTe$_4$, and Fe$_2$WTe$_4$ (see Figs.~\ref{fig2}(a) and \ref{fig2}(b)). The resulting -pCOHP curves reveal that the bonding states below the Fermi level are predominantly occupied, whereas the corresponding anti-bonding states remain largely vacant. This electronic configuration signifies strong and stable bonding interactions.
 
These monolayer materials contain Fe atoms, which possess intrinsic magnetic moments due to their partially filled 3$d$ orbitals. To determine the magnetic ground state, we systematically compared the total energies of two representative magnetic configurations—ferromagnetic (FM) and altermagnetic (AM) [see Fig.~\ref{fig3}(a) for the AM structure]. Our calculations show that all compounds energetically prefer the AM configuration [see Fig.~\ref{fig3}(a)], with the corresponding energy differences between FM and AM states summarized in Table~\ref{table1}. Notably, these energy differences are smaller than those reported in previous work~\cite{li2025ferrovalley}. This discrepancy primarily stems from the different values of the Hubbard \( U \) parameter employed: our study uses \( U = 4\ \text{eV} \), whereas the previous work adopted \( U = 2\ \text{eV} \).

In this ground state, the magnetic moments are mainly localized on the Fe sites, with each Fe atom carrying a moment of approximately $3~\mu_B$. In the absence of SOC, the symmetry of the AM state is described by the spin space group (SSG)~\cite{xiao2024spin,chen2024enumeration,jiang2024enumeration}. Specifically, the systems exhibit symmetry operations $\{C_{2}||M_{xy}\}$ and $\{C_{2}||S_{4z}^+\}$, identifying monolayer Fe$_2$MoTe$_4$ and Fe$_2$WTe$_4$ as $d$-wave altermagnetic materials.

\begin{table*}[htb]
	\caption{Calculated properties of Fe$_2$MoS$_4$, Fe$_2$MoSe$_4$, Fe$_2$MoTe$_4$, and Fe$_2$WTe$_4$, including the optimized lattice constant $a$ (in \AA), the energy difference between altermagnetic (AM) and ferromagnetic (FM) configurations $\Delta E_{\mathrm{AM-FM}}$ (in eV per primitive cell), the exchange parameter J$_1$ (in meV), and the MAE with SOC included: the energy difference between the [001] and [100] directions ($\Delta E_{\mathrm{001-100}}$), and between the [001] and [110] directions ($\Delta E_{\mathrm{001-110}}$), both given in $\mu$eV per primitive cell.}
	\begin{ruledtabular}
			\begin{tabular}{cccccc}
	Systems  & $a$ & $\Delta E_{\mathrm{AM-FM}}$ (eV) & J$_1$(meV)  & $\Delta E_{\mathrm{001-100}}$ ($\mu$eV) & $\Delta E_{\mathrm{001-110}}$ ($\mu$eV) \\
	\hline
	Fe$_2$MoS$_4$   & 5.547    & $-0.05570$   & $-6.962$ & $-248.41$  & $-248.42$  \\
	Fe$_2$MoSe$_4$  & 5.698    & $-0.02846$   & $-3.558$ & $-106.64$  & $-106.59$  \\
	Fe$_2$MoTe$_4$  & 5.935    & $-0.01595$   & $-1.993$ & $-31.36$   & $-31.42$   \\
	Fe$_2$WTe$_4$   & 5.925    & $-0.01251$   & $-1.563$ & $-129.46$  & $-127.42$  \\
\end{tabular}\label{table1}
	\end{ruledtabular}
\end{table*}

	
\subsection{Electronic band and valley structures}
We begin by examining the electronic band structures of monolayer Fe$_2$MoTe$_4$ and Fe$_2$WTe$_4$ in the absence of SOC. The calculated band structures and projected density of states (PDOS) are shown in Figs.~\ref{fig3}(b) and ~\ref{fig3}(c). Both materials exhibit clear spin splitting, and are identified as indirect band gap semiconductors. The spin splitting in the valence bands at both the X and Y points is about 0.71 eV for Fe$_2$MoTe$_4$ and 0.70 eV for Fe$_2$WTe$_4$, which are larger than those in Fe$_2$MoS$_4$ and Fe$_2$MoSe$_4$. Specifically, the spin splittings at the X and Y points are approximately 0.36 eV for Fe$_2$MoS$_4$ and 0.52 eV for Fe$_2$MoSe$_4$, as shown in the SM~\cite{SM}. The conduction band minimum (CBM) is located at the $\Gamma$ point, while the valence band maximum (VBM) appears at the X and Y points. The corresponding global band gaps are 0.666 eV for Fe$_2$MoTe$_4$ and 0.909 eV for Fe$_2$WTe$_4$. The PDOS shows that the low-energy states in the valence band are mainly derived from Fe $d$ orbitals. A noteworthy feature in the band structures is the emergence of a valley degree of freedom. Both materials display well-defined valleys at the X and Y points of the valence band, which are TRIM. Although these valleys are energetically degenerate, they host opposite spin polarization due to the $\{C_{2}||S_{4z}^+\}$ symmetry. Additionally, all bands along the $\Gamma$–M path remain spin-degenerate, a behavior protected by the $\{C_{2}||M_{xy}\}$ symmetry.

	\begin{figure*}[htb]
		\includegraphics[width=17cm]{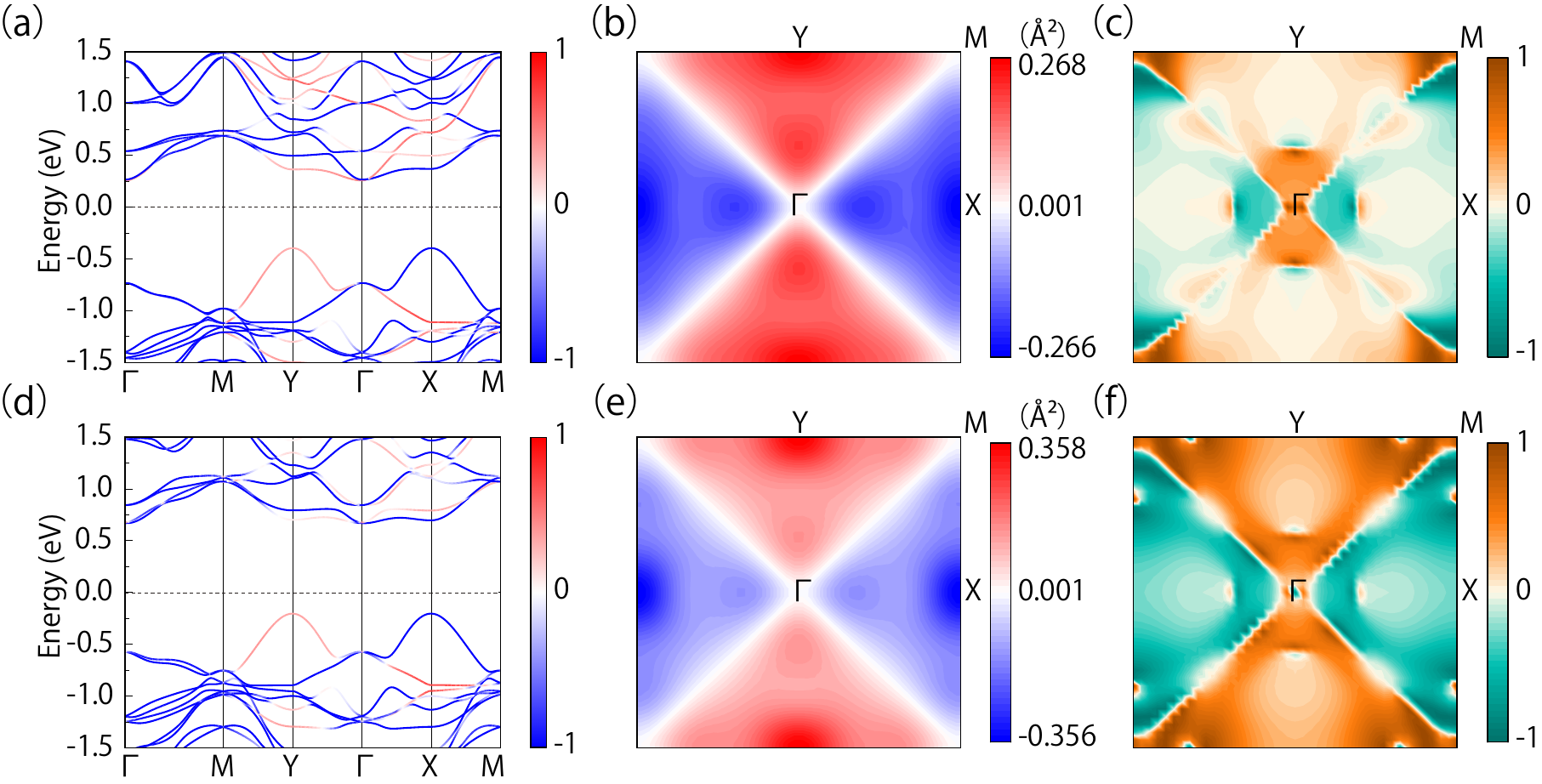}
		\caption{Band structure with the spin projection $s_z$ of monolayer (a) Fe$_2$MoTe$_4$ and (d) Fe$_2$WTe$_4$. Distribution of Berry curvature summed for all valance bands for monolayer (b) Fe$_2$MoTe$_4$ and (e) Fe$_2$WTe$_4$. The circular polarization calculated for transitions between the highest valence band and lowest conduction band for (c) Fe$_2$MoTe$_4$ and (f) Fe$_2$WTe$_4$.}
		\label{fig4}
	\end{figure*}
		
\subsection{Valley-contrasting berry curvature and circular dichroism}
Next, we investigate the electronic band structures of monolayer Fe$_2$MoTe$_4$ and Fe$_2$WTe$_4$ with SOC included. Upon introducing SOC, the symmetry of these monolayers becomes dependent on the orientation of the magnetic moments. To determine the Néel vector direction, we calculate the magnetocrystalline anisotropy energy (MAE) by aligning the magnetization along the [001], [100], and [110] directions and comparing the total energies. The computed energy differences are provided in Table~\ref{table1}. Among the considered directions, the [001] axis is identified as the magnetic easy axis. Due to the different Hubbard $U$ parameters employed, the calculated MAEs are also smaller than those reported in previous work~\cite{li2025ferrovalley}.
The SOC-included band structures of Fe$_2$MoTe$_4$ and Fe$_2$WTe$4$ are shown in Figs.~\ref{fig4}(a) and \ref{fig4}(d), respectively. In these plots, the red and blue colors indicate the spin polarization $\left\langle n\bm{k}\left|\hat{s}_{z}\right| n\bm{k}\right\rangle$ of the eigenstate $|n\bm{k}\rangle$. Both materials retain their nature as indirect band gap semiconductors, with global band gaps of 0.653 eV for Fe$_2$MoTe$_4$ and 0.873 eV for Fe$_2$WTe$_4$. These values are only slightly reduced compared to the SOC-free case. Notably, the previously degenerate bands along the $\Gamma$–M path are now split due to the breaking of the $\{C_{2}||M_{xy}\}$ symmetry induced by SOC.

The valleys in these materials exhibit rich physical phenomena, notably Berry curvature and optical circular dichroism. We begin by discussing the Berry curvature. In 2D systems, the Berry curvature has only one nonzero component—along the out-of-plane ($z$) direction—acting as a pseudoscalar. For a given Bloch state $|n\bm k\rangle$, the Berry curvature is defined as
\begin{equation}
	\Omega_{n\bm k}=-2 \operatorname{Im} \sum_{n'\neq n} \frac{\left\langle n \bm{k}\left|v_{x}\right| n' \bm{k}\right\rangle\left\langle n' \bm{k}\left|v_{y}\right| n \bm{k}\right\rangle}{(\omega_{n^{\prime}}-\omega_{n})^{2}},
\end{equation}
where $v_{x/y}$ are the velocity operators, and $E_n=\hbar\omega_{n}$ is the energy of the state $|n\bm k\rangle$. The total Berry curvature summed over all occupied valence bands, $\Omega(\bm k)=\sum_{n\in occ.} \Omega_{n\bm k}$, is plotted in Figs.~\ref{fig4}(b) and ~\ref{fig4}(e) for monolayer Fe$_2$MoTe$_4$ and Fe$_2$WTe$_4$, respectively. The distribution shows sharp peaks at the X and Y valleys, with opposite signs, reflecting the valley-contrasting nature of the Berry curvature. This feature is consistent with previous reports~\cite{li2025ferrovalley}. This nonzero Berry curvature leads to anomalous Hall transport: an in-plane electric field induces a transverse carrier velocity even without an external magnetic field. Within the semiclassical picture~\cite{xiao2010berry}, the Berry curvature contributes an anomalous velocity term proportional to $\propto \bm E\times\bm{\Omega}$, directly affecting carrier dynamics. The opposite Berry curvature at the X and Y valleys also gives rise to a valley Hall effect, enabling valley-selective transverse responses under an applied in-plane electric field.

The presence of Berry curvature also signals a chiral character of the valleys, which manifests in optical circular dichroism\cite{yao2008valley}. Due to the vertical and spin-conserving nature of optical transitions, excitations occur only between bands within the same spin channel, allowing independent analysis of each spin sector. The circular polarization for interband transitions is defined as
\begin{equation}  
	\eta = \frac{|\mathcal{M}_{+}|^{2} - |\mathcal{M}_{-}|^{2}}{|\mathcal{M}_{+}|^{2} + |\mathcal{M}_{-}|^{2}},  
\end{equation}  
where \(\mathcal{M}_{\pm} = \frac{1}{\sqrt{2}}(\mathcal{M}_x \pm i\mathcal{M}_y)\) are the optical matrix elements corresponding to right- and left-handed circularly polarized light  ($\sigma_{\pm}$), and \(\mathcal{M}_{x/y} = m_e \langle c\bm{k}|v_{x/y}|v\bm{k}\rangle\), with \(m_e\) being the free electron mass. 
Our first-principles calculations yield the distribution of the circular polarization 
\(\eta\) across the BZ for transitions between the highest valence band and lowest conduction band, as shown in Figs.~\ref{fig4}(c) and (f). Remarkably, the dominant optical transitions occur at the X and Y valleys, where 
\(\eta\) takes opposite signs—clearly illustrating valley-dependent optical selection rules and reinforcing the valley-contrasting chiral nature of these systems.

\subsection{Strain-induced valley polarization and piezomagnetism}
	\begin{figure*}[htb]
	\includegraphics[width=17.6cm]{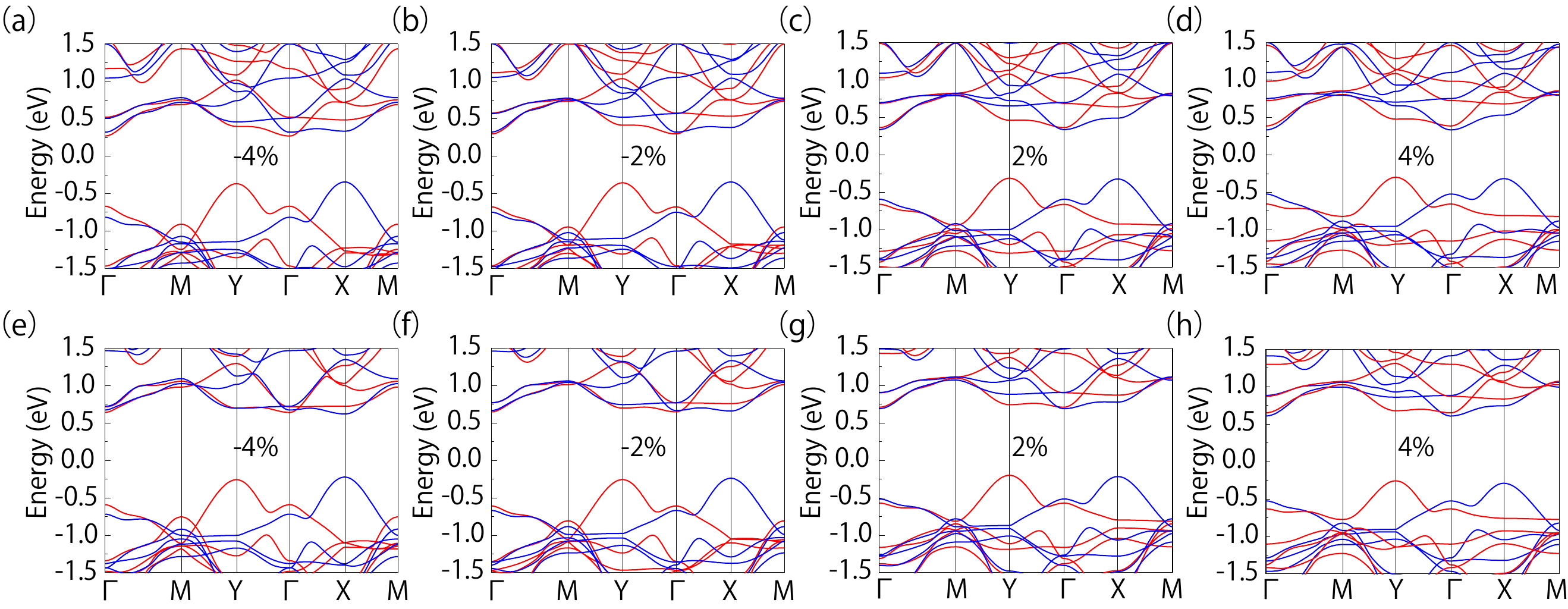}
	\caption{(a)–(d) Band structure evolution under different uniaxial strain along the $a$ direction of monolayer Fe$_2$MoTe$_4$. (e)–(h) Band structure evolution under different uniaxial strain along the $a$ direction of monolayer Fe$_2$WTe$_4$. The SOC is not included.}
	\label{fig5}
\end{figure*}
\begin{figure}[htb]
	\includegraphics[width=8.5cm]{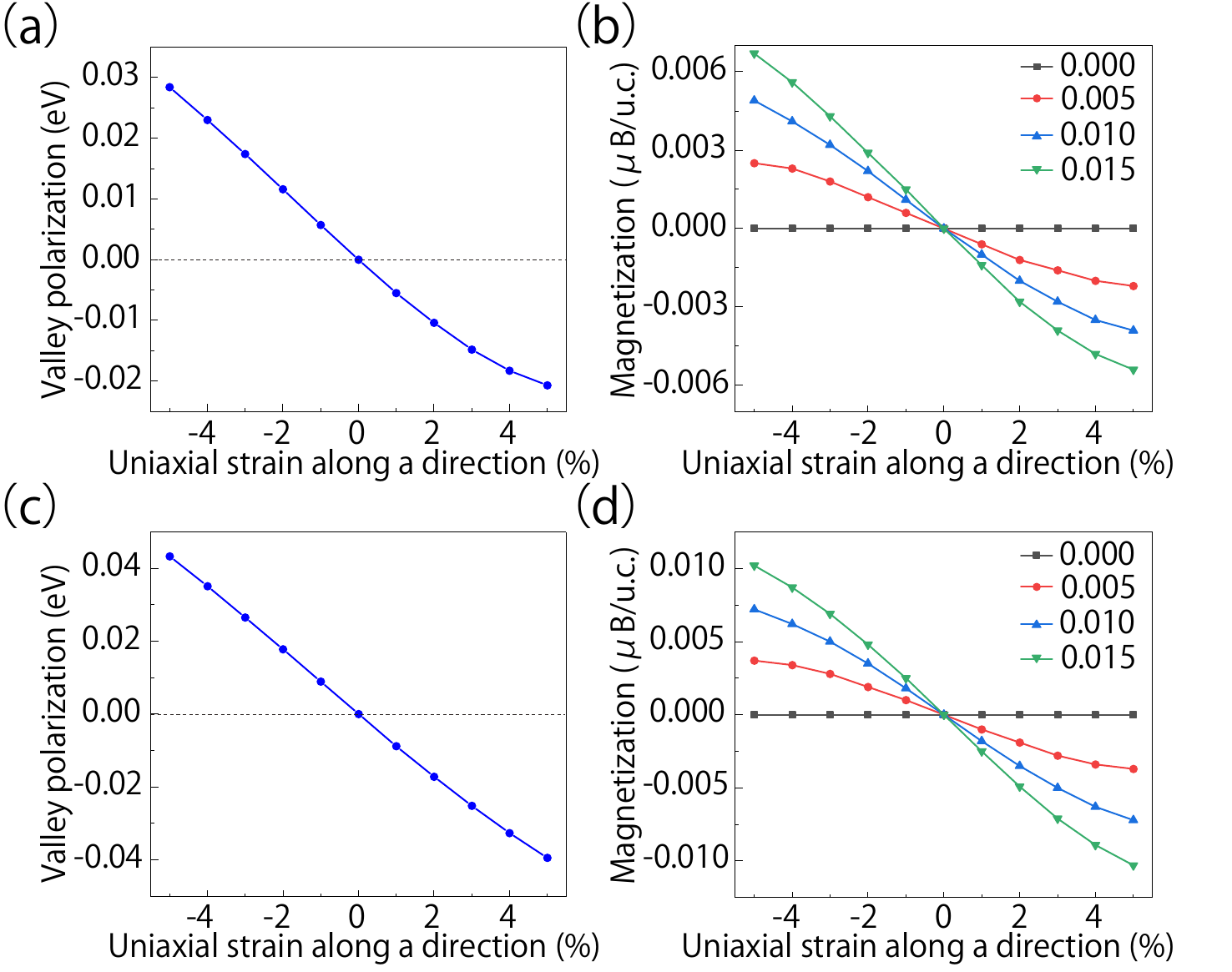}
	\caption{ The strain-controlled valley polarization generated at the valence band of monolayer (a) Fe$_2$MoTe$_4$ and (c) Fe$_2$WTe$_4$. The net magnetization for different uniaxial strains along
		$a$ direction with various hole doping amounts for monolayers (b) Fe$_2$MoTe$_4$ and (d) Fe$_2$WTe$_4$. }
	\label{fig6}
\end{figure}

Through symmetry analysis, the valley degeneracy at the high-symmetry X and Y points in the absence of SOC is protected by the combined $\{C_{2}||M_{xy}\}$ symmetries. When uniaxial strain is applied, these symmetries are broken, leading to the lifting of band degeneracy at the X and Y valleys. The evolution of the band structures of monolayer Fe$_2$MoTe$_4$ and Fe$_2$WTe$_4$ under various uniaxial strains along the $a$ direction is shown in Fig.\ref{fig5}. Under tensile strain, the valence band at the Y point shifts upward, while the one at the X point shifts downward, inducing valley polarization. Conversely, under compressive strain, the valence band at the Y point moves downward, whereas the X-point valence band shifts upward, again resulting in valley polarization. Once the valley energies $E(X)$ and $E(Y)$ become non-degenerate, the degree of valley polarization can be quantified by the energy difference $P = E(X) - E(Y)$. The evolution of this energy difference with respect to the applied uniaxial strain is illustrated in Figs.\ref{fig6}(a) and~\ref{fig6}(c). The influence of strain on the band structure and valleys is similar to that reported in previous studies~\cite{li2025ferrovalley}.

The strain-induced valley polarization provides a promising mechanism for generating net magnetization in these monolayer materials. The net magnetization is given by
$M = \int_{-\infty}^{E_f(n)} [\rho^\uparrow(\epsilon) - \rho^\downarrow(\epsilon)]\, dE,$
where $E_f$ is the Fermi level at a given carrier concentration $n$, and $\rho^{\uparrow(\downarrow)}$ represents the spin-up (down) density of states, both of which depend on the applied strain $\epsilon$. As illustrated in Figs.~\ref{fig6}(b) and~\ref{fig6}(d), undoped monolayers of Fe$_2$MoTe$_4$ and Fe$_2$WTe$_4$ exhibit no net magnetization, even under strain. However, with finite doping, net magnetization emerges and increases with the magnitude of uniaxial strain. Interestingly, compressive and tensile strains induce magnetization in opposite directions. At a fixed strain level, stronger doping generally enhances the magnetization. Moreover, the magnetization displays an approximately linear dependence on small strains and tends to saturate as the strain increases further.

\subsection{Generation of noncollinear spin currents}
\begin{figure*}[htb]
	\includegraphics[width=17cm]{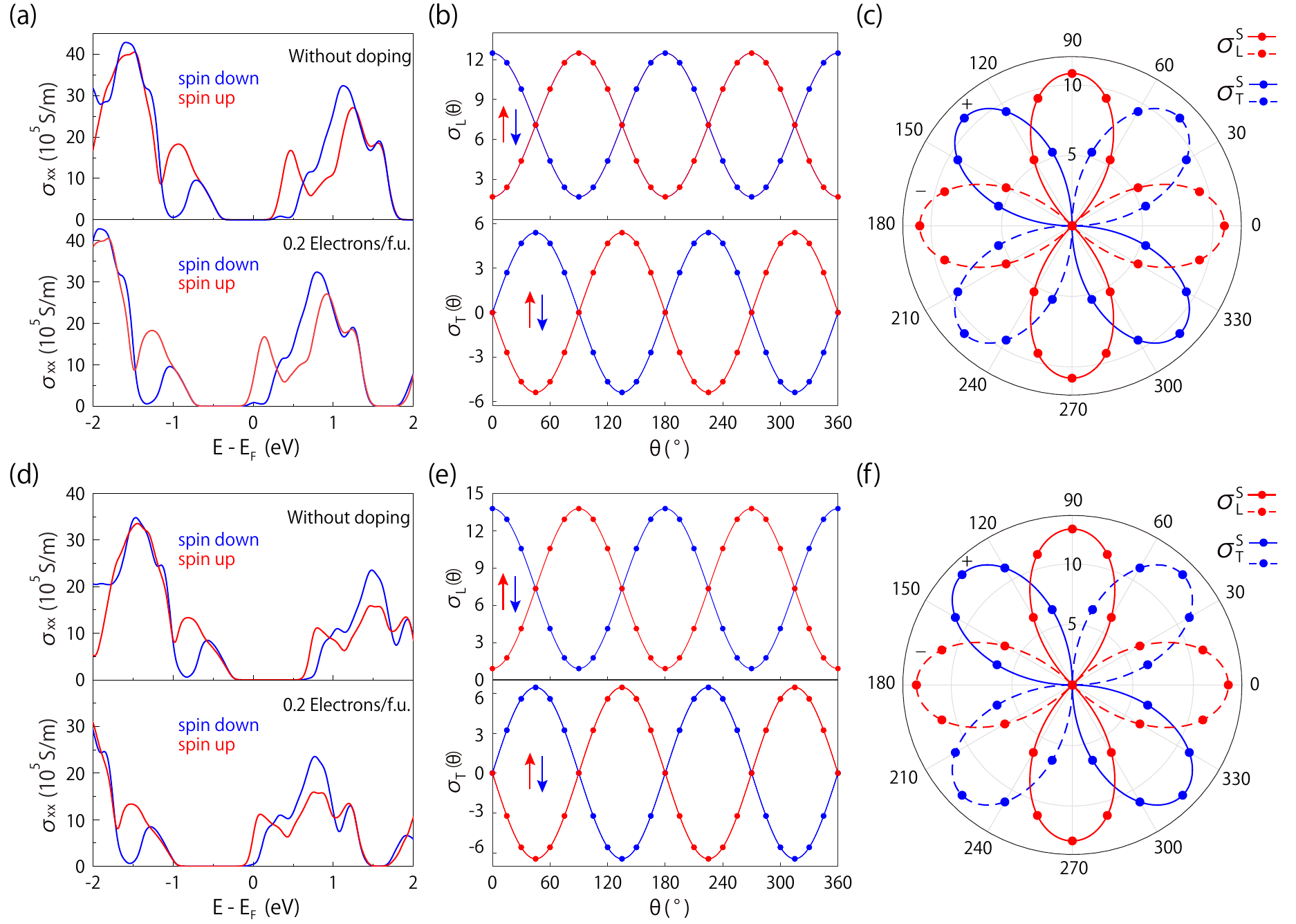}
	\caption{The spin-resolved charge conductivity ($\sigma_{xx}$) of monolayer (a) Fe$_2$MoTe$_4$ and (d) Fe$_2$WTe$_4$ without and with doping with 0.2 electron/f.u. for an electric
		field along the $a$ direction.  Angle dependence of the longitudinal
		($L$) and transverse ($T$) charge conductivity varying with electric field
		direction $\theta$ of monolayer (b) Fe$_2$MoTe$_4$ and (e) Fe$_2$WTe$_4$. Corresponding angle dependence of the longitudinal and transverse spin conductivity of monolayer (c) Fe$_2$MoTe$_4$ and (f) Fe$_2$WTe$_4$. The ``+'' and ``$-$'' symbols represent positive and negative values, respectively. }
		\label{fig7}
\end{figure*}

An altermagnet can drive a spin current when subjected to an in-plane electric field. In the absence of SOC, spin remains a good quantum number, allowing separate definitions and calculations of spin-up and spin-down conductivities($\sigma_0^{\uparrow,\downarrow}$). Figures~\ref{fig7} (a) and (d) display the spin-resolved longitudinal conductivity $\sigma_{xx}$ for monolayer Fe$_2$MoTe$_4$ and Fe$_2$WTe$_4$, both undoped (top panels) and doped with 0.2 electrons per formula unit (bottom panels), under an electric field applied along the $x$ direction. In the undoped case, $\sigma_{xx}$ vanishes at the Fermi level, confirming that both monolayers are insulating. Upon electron doping, a finite $\sigma_{xx}$ emerges around the Fermi level, indicating that the systems become conducting. Because of anisotropic spin splitting in their band structures, the spin-up and spin-down conductivities differ. The spin current can be defined in terms of the spin conductivity as $\sigma_0^{S}=\sigma_0^{\uparrow}-\sigma_0^{\downarrow}$. We further analyze the angular dependence of the longitudinal ($\sigma_L^{\uparrow,\downarrow}$) and transverse ($\sigma_T^{\uparrow,\downarrow}$) conductivities. These satisfy $\sigma _{L }^{\uparrow,\downarrow}(\theta)$ = $\sigma _{0}$ $\mp$ $\sigma _{0}^{\mathrm{S} }$ cos $2\theta$ and  $\sigma_{T}^{\uparrow,\downarrow}(\theta)=\mp\sigma_{0}^{\mathrm{S}}$ sin $2\theta$ and, as shown in Figs.~\ref{fig7} (b) and (e), both vary periodically with the field orientation $\theta$ (period $180^\circ$). The net spin conductivities are given by $\sigma _{L }^{S}(\theta) = \sigma_{L}^{\uparrow}-\sigma_{L}^{\downarrow}=-2\sigma _{0}^{\mathrm{S} }\cos2\theta$ and  $\sigma _{T }^{S}(\theta) = \sigma_{T}^{\uparrow}-\sigma_{T}^{\downarrow}=-2\sigma _{0}^{\mathrm{S} }$ $\sin$ $2\theta$ as plotted in Figs.~\ref{fig7} (c) and (f). Notably, the longitudinal and transverse spin conductivities exhibit a $45^\circ$ phase shift; at $\phi=45^\circ$, $\sigma_L^S$ vanishes, yielding a pure transverse (spin Hall) current.

\bigskip
\section{Discussion and Conclusion}
In this work, we reveal rich valley physics in monolayer Fe$_2$MoX$_4$ (X = S, Se, Te) and Fe$_2$WTe$_4$. It is worth noting that although the conduction band minimum is located at the $\Gamma$ point, this does not affect the characteristics of the valleys at X and Y in the valence band. By introducing hole doping, the Fermi level can be tuned to intersect with the valence band, ensuring that the system’s physical properties are dominated by the X and Y valleys, while the $\Gamma$-point conduction states have little impact on the valley-related phenomena considered here.

In addition, we note the study on the ferrovalley physics of stacked bilayer Fe$_2$MX$_4$-type materials~\cite{li2025ferrovalley}. We also observe that our calculated electronic structures and magnetic properties differ somewhat from those reported in that work. This discrepancy primarily arises from the use of different $U$ values. We have further examined the band structures and magnetic properties under various $U$ values, and the corresponding results are provided in the SM~\cite{SM}.

In conclusion, through first-principles calculations and theoretical analysis, we have systematically explored the valley-dependent electronic properties of 2D altermagnetic iron-based transition metal chalcogenides, Fe$_2$MoX$_4$ (X = S, Se, Te) and Fe$_2$WTe$_4$. Our study reveals that these materials exhibit an altermagnetic ground state and semiconducting behavior, with distinct valley features in the valence band located at the time-reversal-invariant X and Y points. We have examined their Berry curvature and optical circular dichroism, both of which are intimately linked to valley physics. Remarkably, we find that external uniaxial strain can break the valley degeneracy, inducing valley polarization and enabling piezomagnetism under hole doping. Furthermore, we identify sizable noncollinear spin currents in these systems. These findings highlight the rich valley-dependent phenomena in these altermagnetic monolayers and point to their promising potential for applications in valleytronics, spintronics, and nanoelectronics.

\bigskip
\begin{acknowledgements}
	This work is supported by the National Natural Science Foundation of China (Grant No. 12204378) and the Key Program of the Natural Science Basic Research Plan of Shaanxi Province (Grant No. 2025JC-QYCX-007).
\end{acknowledgements}

%

\end{document}